\documentclass[reprint,
 amsmath,amssymb,
 aps,
]{revtex4-2}

\usepackage{graphicx}
\usepackage{dcolumn}
\usepackage{bm}

\usepackage{xcolor}

\begin{document}

\preprint{APS/123-QED}

\title{Epidemic modelling requires knowledge of the social network}



\author{Samuel Johnson}
\email{s.johnson.4@bham.ac.uk}
\affiliation{School of Mathematics, University of Birmingham, Edgbaston B15 2TT, United Kingdom, and\\
The Alan Turing Institute, British Library, 96 Euston Rd, London NW1 2DB, United Kingdom}


\begin{abstract}
`Compartmental models' of epidemics are widely used to forecast the effects of communicable diseases such as COVID-19 and to guide policy. Although it has long been known that such processes take place on social networks, the assumption of `random mixing' is usually made, which ignores network structure. However, `super-spreading events'have been found to be power-law distributed, suggesting that the underlying networks may be scale free or at least highly heterogeneous. The random-mixing assumption would then produce an overestimation of the herd-immunity threshold for given $R_0$; and a (more significant) overestimation of $R_0$ itself. These two errors compound each other, and can lead to forecasts greatly overestimating the number of infections.
Moreover, if networks are heterogeneous and change in time, multiple waves of infection can occur, which are not predicted by random mixing.
A simple SIR model
simulated on both Erd\H os-R\'enyi and scale-free networks shows that details of the network structure can be more important than the intrinsic transmissibility of a disease. It is therefore crucial to incorporate network information into standard models of epidemics.
\end{abstract}

\maketitle


\section{Modelling epidemics}

Throughout the recent COVID-19 pandemic, `compartmental models', such as the SIR or SEIR models, were widely used to forecast
the likely number of infections, hospitalisations and deaths from the disease under different scenarios \cite{abou2020compartmental,gnanvi2021reliability,mccabe2021modelling}, particularly as a guide to making `non-pharmaceutical interventions' (NPIs) \cite{ferguson2020impact,vardavas2021modeling}.
However, doubts arose as to their predictive power \cite{ioannidis2020forecasting}.
In the UK, for example, the large waves of infections expected to occur in the absence of NPIs, in both the summer and winter of 2021, failed to materialise \footnote{See ``SPI-M-O: Consensus Statement on COVID-19'', from 15 December 2021, for the advice given to the UK government before the expected Omicron wave: \url{https://assets.publishing.service.gov.uk/government/uploads/system/uploads/attachment_data/file/1042204/S1439_SPI-M-O_Consensus_Statement.pdf}}.

While these models can take account of many details of how the specific disease spreads, they usually make the assumption of `random mixing': that any individual can infect any other \cite{keeling2005networks}. However, people are in fact connected according to a social network \cite{Newman_rev}. It has been known for decades that network topology has important effects on spreading processes \cite{moore2000epidemics,Newman_rev,pastor2001epidemic,newman2002spread,keeling2005networks,morita2016six,moore2020predicting}, but in practice it is difficult to gather data on this web of contacts. Moreover, accounting for the network explicitly with agent-based modelling may be computationally prohibitive at the scale of, say, a whole country.
Hence, random mixing -- albeit with a degree of structure captured by the inclusion of different groups of people  -- remains the standard assumption \cite{keeling2020fitting}.

Social networks of various kinds have been found to be highly heterogeneous, in that the degree, $k$, of vertices (i.e. the number of contacts of each person) follows a distribution with a high variance \cite{newman2002random,jackson2007meeting}. Scale-free networks -- in which this distribution is a power law, $p(k)\sim k^{-\alpha}$, with $\alpha$ usually between 2 and 3 -- are an extreme example of this. For instance, a network of sexual contacts was observed to follow this rule with $\alpha\simeq 2.4$ \cite{liljeros2001web}.
And while we don't have detailed information on the network of contacts underlying the spread of respiratory viruses, we do know that, 
in the early stages of an epidemic,
COVID-19 is driven largely by `super-spreading events' (SSEs) \cite{lewis2021superspreading}, with one estimate suggesting that fewer than $10\%$ of infectious individuals accounted for $80\%$ of infections \cite{endo2020estimating}. The importance of SSEs was also shown in the case of SARS \cite{riley2003transmission}. Moreover, Fukui and Furukawa \cite{fukui2020power} found the distribution of these SSEs -- that is, the number of individuals infected on each occasion -- followed power laws in the cases of SARS, MERS and COVID-19. This suggests that the underlying networks have highly heterogeneous degree distributions, which would be consistent with other studies of social networks \cite{newman2002random,jackson2007meeting,liljeros2001web}.

\begin{figure*}[t]
\includegraphics[scale=0.6]{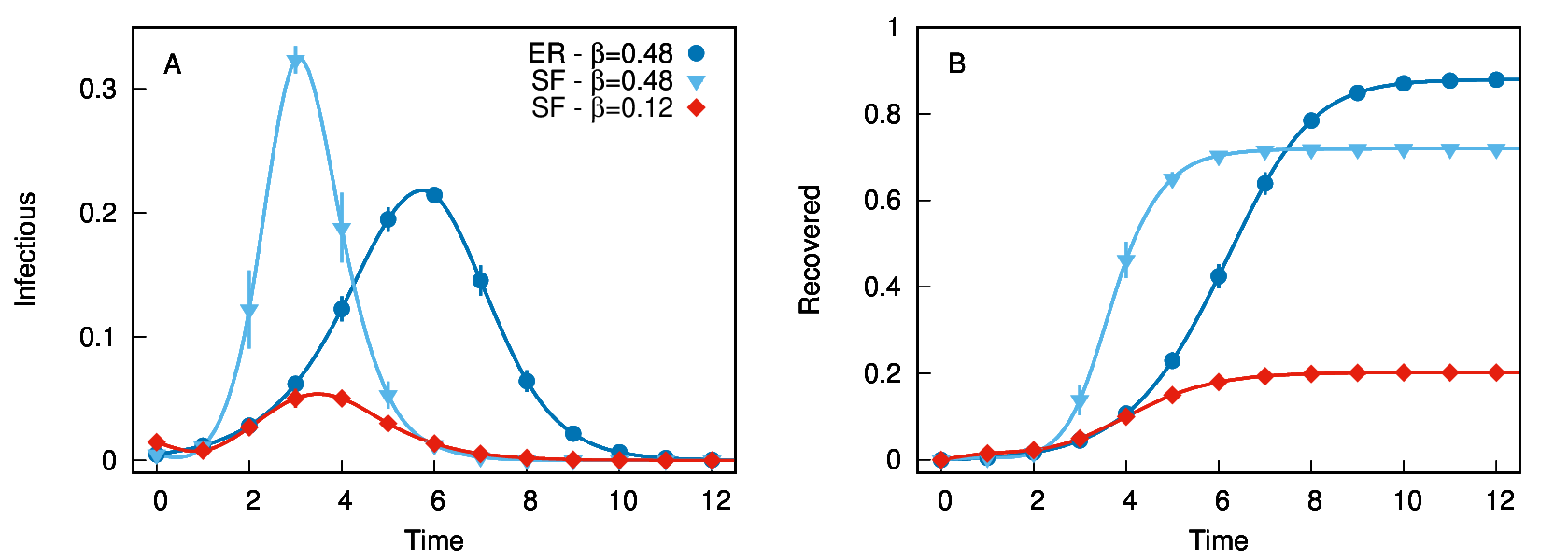}
\caption{\label{fig_1} Time series for the proportions of agents in the Infectious (panel {\bf A}) and Recovered ({\bf B}) states, from the SIR model described in the main text, in three scenarios: Erd\H os-R\'enyi (ER) random graphs and probability of infection $\beta=0.48$ (dark blue circles); scale-free (SF) networks with exponent $\alpha=2.5$ and $\beta=0.48$ (light blue triangles); and SF networks with $\alpha=2.5$ and $\beta=0.12$ (red diamonds). Number of vertices $N=10^4$, mean degree $\langle k\rangle = 5$, averages over $100$ networks in each case, bars represent one standard deviation. At time $t=0$ all agents are Susceptible, except for 150 randomly chosen agents set to Infectious, for the SF network with $\beta=0.12$; or 50 randomly chosen agents for the other two cases (the discrepancy is to showcase the overlapping curves better). Lines (splines) are a guide for the eye.}
\end{figure*}

This letter uses an agent-based version of an SIR model to illustrate how the random-mixing assumption can lead to very large errors in the total number of people predicted to become infected,
 in a given epidemic `wave',
if the network is scale free. This is not just because, for a given basic reproduction number, $R_0$, the `herd immunity threshold' (HIT) is generally lower on a scale-free network;
but, more significantly, because the initial rapid growth in infections in the scale-free case can lead to an overestimation of $R_0$.
As we shall see, the combination of these two effects can produce a random-mixing forecast of over $80\%$ of the population becoming infected, when in fact only $20\%$ are affected before the epidemic dies down naturally.

Conversely, once an epidemic has reached herd immunity, the random-mixing assumption predicts that the population is safe from further waves unless immunity wanes. However, if the networks are scale free and change in time, multiple waves can occur even as individual immunity is maintained.

The COVID-19 pandemic involved multiple waves of infection in countries with quite different levels of stringency in their NPIs. This seems more compatible with a process taking place on time-varying, heterogeneous networks than with the predictions of random-mixing models.

\section{The effect of the network}
Consider a network in which each vertex represents an agent, and edges are contacts which potentially allow for contagion of a transmissible disease.
We will compare here two different topologies: Erd\H{o}s-R\'enyi (ER) networks, in which the edges are placed entirely at random among the vertices \cite{ErdosRenyi}; and scale-free (SF) networks. The latter are constructed by drawing desired degrees from a distribution $p(k)\sim k^{-\alpha}$, and using the `configuration model' to place the edges \cite{Newman_rev}. A `structural cut off' is imposed, such that $k<\sqrt{\langle k\rangle N}$, where $\langle k\rangle$ is the mean degree and $N$ the number of vertices. For the parameters used here -- $\langle k\rangle = 5$ and $N=10^4$ -- the maximum degree in the SF case is therefore 233. This is not unrealistically high for COVID-19 contact networks since some SSEs saw over 100 people apparently infected by a single individual within a few hours. In both cases we will consider undirected networks for simplicity, although directionality has been found to have an important influence on spreading processes \cite{klaise2016neurons}.
The random-mixing assumption is a good mean-field description of the Erd\H{o}s-R\'enyi case. However, as discussed, the scale-free network may be a better model for a real web of social contacts.

\begin{figure*}[t]
\includegraphics[scale=0.6]{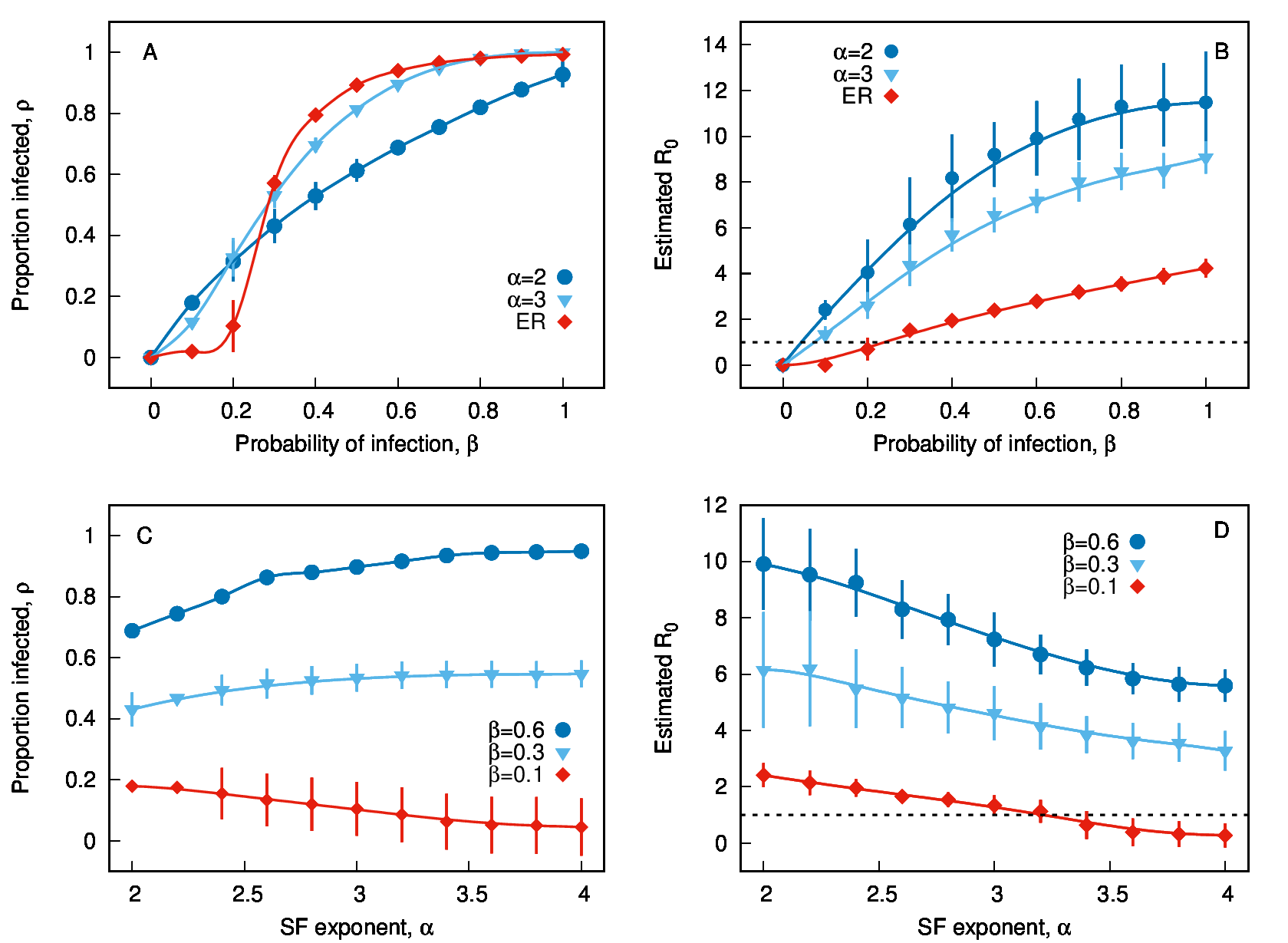}
\caption{\label{fig_hetero} Proportion of agents ever infected, $\rho$, against probability of infection, $\beta$, for SF networks with $\alpha=2$ (dark blue circles) and $\alpha=3$ (light blue triangles), and for ER random graphs (red diamonds) (Panel {\bf A}). Estimated value of basic reproduction number, $R_0^e$, from Eq. (\ref{eq_R0e}) against $\beta$ on the same networks ({\bf B}).
Proportion infected, $\rho$, against SF exponent $\alpha$ for infection probability $\beta=0.6$ (dark blue circles), $\beta=0.3$ (light blue triangles) and $\beta=0.1$ (red diamonds) ({\bf C}). And $R_0^e$ against $\alpha$ for the same values of $\beta$ ({\bf D}). All agents are initially Susceptible except for 50 randomly chosen to be set to Infectious. 
All other parameters as in Fig. \ref{fig_1}.
}
\end{figure*}

The epidemic is described
by the following model. Every agent $v_i$ has a state 
$z_i(t)$
at discrete time $t$, which can take one of three values: $S$, $I$ or $R$ (Susceptible, Infectious or Recovered).
If there is an edge from $v_i$ to $v_j$, and if $z_i(t)=I$ and $z_j(t)=S$, then with probability $\beta$ we set $z_j(t+1)=I$ (i.e. $v_j$ is infected by $v_i$).
If $z_i(t)=I$, then $z_i(t+\tau)=R$, for all $\tau\geq 1$ (i.e. every agent recovers after one time step, and thereafter cannot change state, as though either immune or deceased).
Agents are updated in parallel at every time step.

This is a very simple version of an SIR model, with no allowance made for heterogeneity in transmission times, infectiousness or other features, nor for different categories of agents, such as asymptomatic individuals, children, etc.
Moreover, parallel updating is not always a good approximation for a continuous-time process \cite{fennell2016limitations}, which would be simulated more realistically with a Gillespie algorithm \cite{cota2017optimized}. The purpose of this model here
is merely to highlight
how
knowledge of the network is crucial even in the simplest of settings.

\begin{figure}[t]
\includegraphics[scale=0.65]{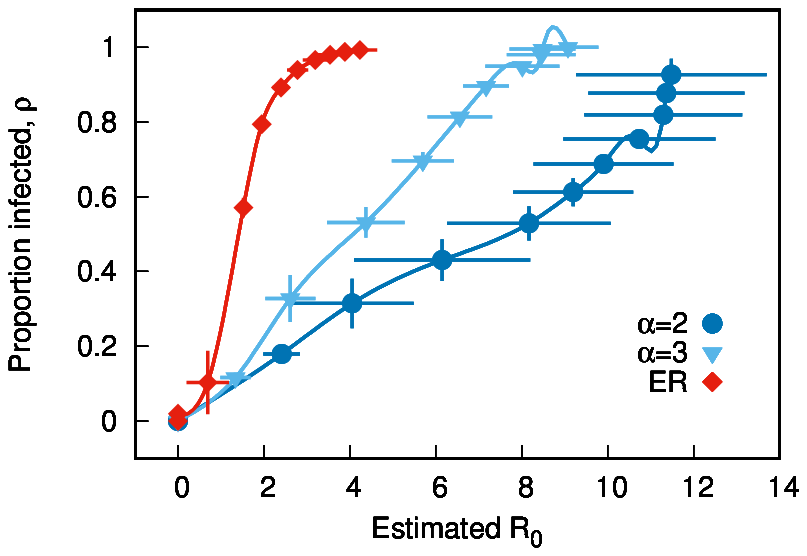}
\caption{\label{fig_RI} Proportion of agents ever infected, $\rho$, against estimated basic reproduction number, $R_0^e$, from Eq. (\ref{eq_R0e}) for SF networks with exponent $\alpha=2$ (dark blue circles) and $\alpha=3$ (light blue triangles), and for ER random graphs (red diamonds). Different values for the same network correspond to the different values of $\beta$ used in Fig. \ref{fig_hetero} {\bf A} and {\bf B}. All other parameters as in Fig. \ref{fig_hetero}.
}
\end{figure}

Consider the situation where initially all agents are Susceptible except for one randomly chosen agent, which is made Infectious at time $t=0$. If the mean degree of the network is $\langle k\rangle$, at $t=1$ the expected number of Infectious 
agents
will be $\langle k\rangle \beta$, so the basic reproduction number will be
\begin{equation}
R_0=\langle k\rangle \beta.    
\label{eq_R0}
\end{equation}
The expected mean degree of the newly Infectious agents, however, is not $\langle k\rangle$, but $\langle k^2\rangle/\langle k\rangle$, where $\langle k^2\rangle$ is the second moment of the degree distribution. (This is an instance of the `friendship paradox': your friends have more friends than you \cite{jackson2019friendship}). So, taking into account that one of the contacts was the originally Infectious vertex, the effective reproduction number, $R_t$, at the next time step 
($t=1$)
is
\begin{equation}
R_1=\left(\frac{\langle k^2\rangle}{\langle k\rangle}-1\right) \beta = \left(\frac{\sigma^2}{\langle k\rangle}+\langle k\rangle-1\right) \beta,
\label{eq_R1}
\end{equation}
where $\sigma^2$ is the variance of the degree distribution, $p(k)$. If the network is an Erd\H{o}s-R\'enyi random graph, this is a Poisson distribution, so $\sigma^2=\langle k\rangle$ and $R_1=R_0$. However, if degrees are distributed more heterogeneously, as in a scale-free network,
then $R_1>R_0$ \cite{Newman_rev}. In other words, the epidemic accelerates as it reaches more highly connected vertices (hubs).

On the other hand, the epidemic plays the role of a targeted attack on the network: by infecting the hubs early on, it removes edges more rapidly in a more heterogeneous network than in a homogeneous one, with the result that in the heterogeneous case fewer vertices may end up becoming Infectious before the epidemic peters out. This is an instance of a more general effect whereby if susceptibility and infectiousness are both heterogeneously distributed and positively correlated in a population, the `herd immunity threshold' (HIT) (i.e. the proportion of infected people when $R_t$ drops below one) is lower than we would expect from the standard equation HIT$=1-1/R_0$, which follows from the assumption of random mixing \cite{gomes2022individual}.

Figure \ref{fig_1} shows averages over time series for the proportion of the agents which are Infectious (panel {\bf A}) or Recovered ({\bf B}), for three different scenarios. The 
dark blue circles
correspond to Erd\H os-R\'enyi random graphs with
$\langle k\rangle =5$. The infection probability is $\beta=0.48$ so, according to Eq. (\ref{eq_R0}), $R_0=2.4$.
Eventually about $88\%$ of agents become infected. This is fairly close to the prediction of $81\%$ for COVID-19 infections in the UK and the US made in March 2020 by the group led by Prof. Neil Ferguson \cite{ferguson2020impact}, despite the much greater sophistication of their model, for the case in which no NPIs were introduced and based on an estimate of $R_0=2.4$.

The light blue triangles in Fig. \ref{fig_1} are for the same parameter values ($\langle k\rangle =5$ and $\beta=0.48$) but now the networks are scale free, with an exponent $\alpha=2.5$.
The curve now grows significantly faster and peaks at a higher value, yet also falls more quickly, going on eventually to infect a slightly smaller proportion of the population ($72\%$) than in the ER case.

The red diamonds also correspond to SF networks with $\alpha=2.5$, but now $\beta=0.12$. In this case, the curve initially follows a very similar trajectory to the ER network with $\beta=0.48$; but it peaks earlier at a lower value, and goes on to infect only
$20\%$ of the population.

This example serves to illustrate how two different scenarios -- high transmissibility on a homogeneous network, and low transmissibility on a heterogeneous one -- can initially follow very similar epidemic curves, yet go on to have markedly different outcomes.

\section{Mismeasuring $R_0$}
In practice, it is not usually possible to obtain the value of $R_0$ from contact tracing. Rather, scientists estimate this number from the rate at which infections grow in the early stages of the epidemic, together with assumptions about the incubation period and duration of infectiousness \cite{ferguson2020impact,anderson2020reproduction}. For instance, if one assumes that each Infectious individual infects $R_0$ others after a period $\tau$, and the number of Recovered is low enough that one can assume exponential growth, then the number of Infectious individuals at time $t$ is
\begin{equation}
I(t)=I(0) R_0^{t/\tau}. 
\label{eq_It}
\end{equation}

Imagine a group of scientists living in a scale-free world who observed an epidemic growing, in its early stages, as the red diamonds of Fig. \ref{fig_1}. If they assumed random mixing and estimated $R_0$ from Eq. (\ref{eq_It}), they would find that $R_0\simeq 2.4$. Their model, even if quite sophisticated in other ways, may well then predict that the epidemic would evolve similarly to the dark blue circles. Moreover, if NPIs were then imposed, and the curve went on to peak earlier than forecast and well before the expected HIT, it would be natural to assume that $R_t$ had fallen below one thanks to the NPIs. Only when an epidemic were allowed to spread without added NPIs would it become apparent that the model's predictions were significantly wrong. 

Figure \ref{fig_hetero} ({\bf A}) shows the proportion of the population who have been infected after the wave has passed, $\rho=\lim_{t\rightarrow\infty} R(t)$, against $\beta$ for SF networks with $\alpha=2$ and $3$, and for ER networks. At low $\beta$, the epidemic reaches more agents on the SF networks, since the process does not percolate on ER networks for $\beta \langle k\rangle < 1$. However, for larger $\beta$ the epidemic reaches more agents on more homogeneous networks (i.e. the HIT is lower on SF networks \cite{fukui2020power}). 

Figure \ref{fig_hetero} ({\bf B}) shows the `estimated $R_0$', or $R_0^e$. Using Eq. (\ref{eq_It}) and bearing in mind that in this model $\tau=1$, this is defined as
\begin{equation}
R_0^e= \max_t \frac{I(t+1)}{I(t)}.
\label{eq_R0e}
\end{equation}
In other words, $R_0^e$ is akin to the value of $R_0$ that a group of scientists might estimate from observations of the doubling time in the early stages of the epidemic.
For ER networks, which are equivalent to random mixing, $R_0^e$ will be very close to $R_0$, as given by Eq. (\ref{eq_R0}) ($R_0^e\simeq \langle k\rangle \beta$). However, we shall see that, for SF networks, $R_0^e$ can be significantly higher than this value ($R_0^e  > \langle k\rangle \beta$). Thus, estimates of the transmissibility of a disease based on changes in the number of cases can be wrong if the underlying social network is heterogeneous. 

Fig. \ref{fig_hetero} ({\bf C}) shows again the eventual proportion of infected agents, but against $\alpha$ for SF networks, and different values of $\beta$. As $\alpha$ decreases, $\beta$ has less of an effect on the reach of the epidemic -- suggesting that the intrinsic transmissibility of a disease is less significant if the network is highly heterogeneous.

Fig. \ref{fig_hetero} ({\bf D}) shows $R_0^e$ against $\alpha$ for SF networks. The estimated reproduction number is always greater the more heterogeneous the network, and in the $\beta=0.1$ case the value of $\alpha$ can even determine whether $R_0^e$ is greater or less than one.

Another way of viewing these results is to plot $\rho$ against $R_0^e$, as in Fig. \ref{fig_RI}. On the ER networks, $\rho$ is very sensitive to $R_0^e$, as in random-mixing models.  But as degree heterogeneity increases, this sensitivity decreases. For example, on the SF networks with $\alpha=2$, there is a range for which a doubling in $R_0^e$ leads to barely a $20\%$ increase in the proportion infected. Hence, if the network is highly heterogeneous, the estimated $R_0$ is very sensitive to $\beta$, yet the number of people who will become infected is not. In other words, it becomes more important to gain knowledge about the network than about the intrinsic transmissibility of the disease.

\begin{figure*}[t]
\includegraphics[scale=0.65]{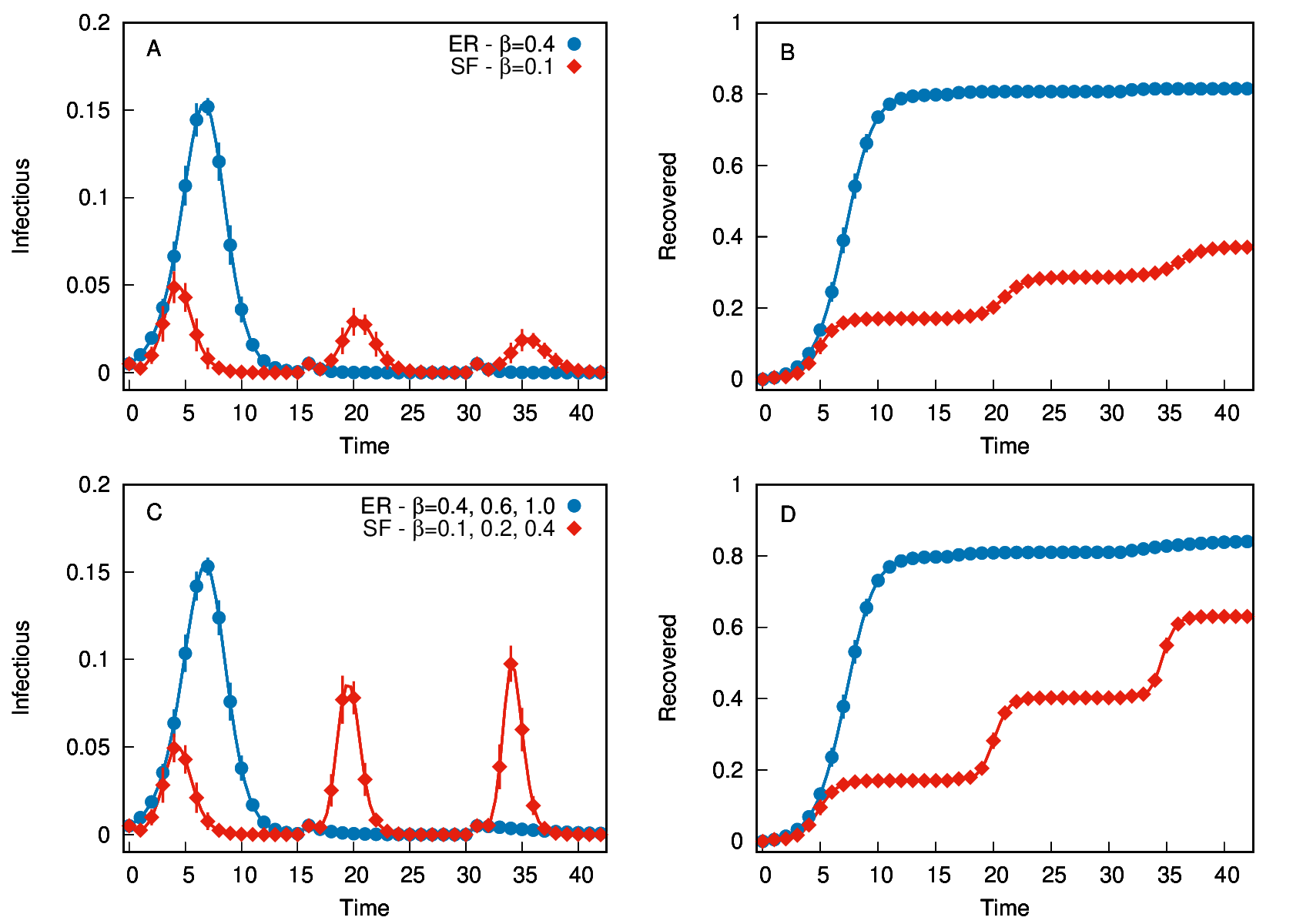}
\caption{\label{fig_waves} 
Time series for proportions of agents in the Infectious (panels {\bf A} and {\bf C}) and Recovered ({\bf B} and {\bf D}) states for ER random graphs (blue circles) and SF networks with exponent $\alpha=2.2$ (red diamonds). At time $t=0$ all agents are Susceptible, except for 50 randomly chosen agents set to Infectious. At times $t=15$ and $t=30$, the networks are replaced with new ones, randomly generated with the same network parameters; and 50 randomly chosen Susceptible agents are set to Infectious. Panels {\bf A} and {\bf B}: Transmissibility is constant at $\beta=0.4$ in the ER case and $\beta=0.1$ in the SF case. Panels {\bf C} and {\bf D}: Transmissibility is increased at times $t=15$ and $t=30$. In the ER case, $\beta=0.4$ until $t=15$, $\beta=0.6$ until $t=30$, and $\beta=1$ thereafter. In the SF case, $\beta=0.1$ until $t=15$, $\beta=0.2$ until $t=30$, and $\beta=0.4$ thereafter. All other parameters as in Fig. \ref{fig_1}.
}
\end{figure*}

\section{Multiple waves}

Once an epidemic has petered out naturally, it is often assumed that herd immunity must have been achieved, and the population is no longer vulnerable unless immunity wanes or transmissibility increases significantly. However, when the HIT is low thanks to the heterogeneity of the social network, a large pool of susceptible individuals may still remain even after a first `wave' of infection. As long as the structure of the network is unchanged, the population will indeed have herd immunity. But if this structure is altered the population may become vulnerable to subsequent waves of infection.

Figure \ref{fig_waves} compares time series for ER and SF networks, as in Figure \ref{fig_1}, but now at times $t=15$ and $t=30$ the network structure is replaced with a new one, and the epidemic is re-seeded by switching a small number of Susceptible agents to Infected (all Recovered agents remain Recovered). Figures \ref{fig_waves} {\bf A} and {\bf B} show the proportions of Infectious and Recovered agents, respectively. Once the epidemic has died down in the ER case, replacing the network with a new version and re-seeding the epidemic has virtually no effect, since there are insufficient remaining Susceptible agents for a new wave to occur. However, in the SF case, a new wave is seen every time the network is changed - albeit with each wave being smaller than the last.
In panels {\bf A} and {\bf B} the transmissibility is constant ($\beta=0.4$ and $0.1$ for the ER and SF networks, respectively). Figures \ref{fig_waves} {\bf C} and {\bf D}, however, show time series for which, in addition to the network structure being changed, the transmissibility is increased to $\beta=0.6$ (at $t=15$) and $\beta=1$ (at $t=30$) for the ER networks; and to $\beta=0.2$ (at $t=15$) and $\beta=0.4$ (at $t=30$) for the SF networks. In the ER case there are still no more waves of infection. However, in the SF case there are now subsequent waves of increasing size.

In real life, should we expect the social networks behind epidemics to change? Certain connections may be quite stable, such as those between work colleagues, while others are transitory, say among people who happen to be attending the same event. The COVID-19 pandemic involved several waves of infection, something variously attributed to more infectious variants of the virus, changing NPIs or waning immunity. However, Figure \ref{fig_waves} shows that an underlying network which is both heterogeneous and time varying is enough to produce several waves, even when previous ones died down naturally.

\section{Conclusion}

While we may not have detailed information on the web of contacts underlying a process such as a COVID-19 epidemic, we know that social networks of various kinds have been found to be highly heterogeneous \cite{Newman_rev}, and that super-spreading events for this and similar diseases appear to be power-law distributed \cite{fukui2020power}. A heterogeneous topology, such as a scale-free network, may therefore be a better null model than the assumption of random mixing.

The epidemic model used here is very simple and devoid of any realistic parameters. But there is no obvious reason to believe that the greater sophistication of the compartmental models often used to guide public health policy would annul the effects reported here. In any case, perhaps this could be explored by implementing versions of such models on networks.
Another caveat is that in this model recovered agents can never again become infected. In reality, we know that diseases such as COVID-19 can re-infect, either because of waning immunity or new variants.
Multiple waves of infection are thus often attributed to changing levels of individual immunity. However, we have seen that a changing network structure, if heterogeneous, can also lead to multiple waves even when individual immunity is maintained.

If these results do carry over to more realistic scenarios, then it is crucial to gather data on the networks of contacts on which epidemics play out, and to adapt existing compartmental models either to correct for network topology, or to take it into account explicitly.
It may be the case that estimating $\langle k^2 \rangle/\langle k \rangle$ in a social network is in fact easier than inferring the mean degree, since methods such as respondent-driven sampling suffer from a bias towards more highly connected individuals \cite{mills2014errors}. 
Further research is also needed to elucidate to what extent social networks change in time and how this affects epidemics \cite{karsai2014time}.

In any case, the effects reported here suggest that: a) each `wave' of a disease such as COVID-19 may infect fewer people than we would otherwise assume, even in the absence of NPIs, thanks to network heterogeneity;
b) if networks are heterogeneous and change in time, this can lead to multiple waves of infection that would not be predicted by random-mixing models; and c)
NPIs focused on avoiding super-spreading events are likely to be particularly efficacious at suppressing the epidemic.

Other network properties -- such as 
efficiency
\cite{latora2001efficient},
assortativity \cite{johnson2010entropic}, directionality \cite{johnson2020digraphs} or spatial
aspects \cite{barthelemy2011spatial} -- may also be as relevant as degree heterogeneity. 
A well-defined community structure, in particular, can have an important effect \cite{lieberthal2023epidemic}.
Ultimately, epidemics are yet another example of how the architecture of complex systems is fundamental to their dynamical behaviour \cite{pastor2001epidemic,newman2002spread,fukui2020power}.

\begin{acknowledgments}
I am grateful for support from the Alan Turing Institute under EPSRC Grant EP/N510129/1.
\end{acknowledgments}


%

\end{document}